\documentclass[pre,twocolumn]{revtex4}
\usepackage{amssymb}

\input{tcilatex}

\begin{document}

\title[Arching in shaken deposits of granular disks]{Arching in tapped
deposits of hard disks.}
\author{Luis A. Pugnaloni}
\affiliation{Instituto de F\'{\i}sica de L\'{\i}quidos y Sistemas Biol\'{o}gicos,
UNLP-CONICET - cc. 565, 1900 La Plata, Argentina}
\author{Marcos G. Valluzzi}
\affiliation{Instituto de F\'{\i}sica de L\'{\i}quidos y Sistemas Biol\'{o}gicos,
UNLP-CONICET - cc. 565, 1900 La Plata, Argentina}
\author{Lucas G. Valluzzi}
\affiliation{Instituto de F\'{\i}sica de L\'{\i}quidos y Sistemas Biol\'{o}gicos,
UNLP-CONICET - cc. 565, 1900 La Plata, Argentina}

\begin{abstract}
We simulate the tapping of a bed of hard disks in a rectangular box by using
a pseudo dynamic algorithm. In these simulations, arches are unambiguously
defined and we can analyze their properties as a function of the tapping
amplitud. We find that an order--disorder transition occurs within a narrow
range of tapping amplitudes as it has been seen by others. Arches are always
present in the system althought they exhibit regular shapes in the ordered
regime. Interestingly, an increase in the number of arches does not always
correspond to a reduction in the packing fraction. This is in contrast with
what is found in three-dimensional systems.
\end{abstract}

\volumeyear{}
\volumenumber{}
\issuenumber{}
\eid{}
\date{7 January 2006}
\received[Received text]{}
\revised[Revised text]{}
\accepted[Accepted text]{}
\published[Published text]{}
\startpage{1}
\endpage{}
\maketitle

\section{Introduction}

The study of structural properties of assemblies of macroscopic particles
packed in a container has become an important area of granular matter.
Properties such as packing fraction ($\phi $), coordination number ($z$),
and arch distributions depend on the way the packings are created. The
so-called "Chicago experiment" \cite{Nowak1} has shown that a tapped
granular bed can achieve a stationary state where the system characterizing
parameters (in this case $\phi $) depend only on the tapping amplitude. This
simplifies our investigations when we are interested only in the steady
state of the system.

Arching is one of the collective phenomena that many associate to the
appearance of voids in a granular sample that leads to the lowering of $\phi 
$. Moreover, arching is directly related to the reduction of
particle--particle contacts in the assembly, which determines the value of $%
z $ \cite{Pugnaloni1,Pugnaloni2}. Arch formation is crucial in the jamming
of granular flows \cite{To1,To2,Zuriguel1,Zuriguel2} and it has also been
proposed as a mechanism for size segregation \cite{Durand1,Durand2}.

Two-dimensional (2D) granular systems merit attention due to the multiple
applications in industry and urban life (pills, bottles, etc. on a conveyor
belt \cite{Mattone1}, traffic jams \cite{Helbing2}, and crowd control \cite%
{Helbing1}) and also due to the possibility of testing theoretical models
that are particularly simple to treat in 2D.

Simulation studies of the shaking of 2D granular beds have been carried out
by others in the past (see for example Ref. \cite{Vidales1}). However, these
pseudo dynamic simulations do not consider simultaneous deposition of the
grains. This prevents arching during deposition, which is a main ingredient
in granular systems. On the other hand, simulations through realistic
granular dynamics have the drawback that there is not a clear criterion to
decide when the system is fully deposited, since kinetic energy does not
vanishes completely in these simulations. Besides, identifying which
particles support a given particle in these type of simulations may prove
rather complex.

In this work we study the formation of arches in a 2D assembly of hard disks
by using a pseudo dynamic simulation of the tapping and simultaneous
deposition of inelastic rough particles. We follow the changes in $\phi $, $%
z $, and arch properties as a function of the tapping amplitude. Also, an
annealed tapping similar to the "Chicago experiment" is carried out on the
system.

\section{Arching and coordination number}

Arches are multi-particle structures formed during simultaneous
(non-sequential) deposition of an assembly of granular particles. Particles
in an arch support each other so that the whole set remains stable against
gravity (or the driving force that promotes deposition) \cite{Pugnaloni2}.

In a two-dimensional bed of convex particles constructed by sequential
deposition, the mean coordination number $\langle z\rangle $ is 4: each
newly deposited particle adds two contacts ---which stabilize the
particle--- to the system. If particles are deposited non-sequentially,
mutual contacts are created where two particles share a stabilizing contact.
This diminishes in 1 ---with respect to a sequential deposition--- the total
number of contacts in the system and so the value of $\langle z\rangle $
drops. Arches in 2D are string-like: any arch has two end-particles that
share a single mutual contact and a variable number of center-particles that
share two mutual contacts each. If $n(s)$ is the number of arches consisting
of $s$ particles in a system of $N$ particles, there are

\begin{equation}
p_{1}=2\sum_{s=2}^{N}n(s)  \label{eq1}
\end{equation}%
end-particles and

\begin{equation}
p_{2}=\sum_{s=3}^{N}(s-2)n(s)  \label{eq2}
\end{equation}%
center-particles in the assembly. Here, $n(s=1)$ corresponds to the number
of particles that do not belong to any arch. Therefore, the mean
coordination number can be obtained as

\begin{equation}
\langle z\rangle _{\text{support}}=4-\frac{p_{1}}{N}-2\frac{p_{2}}{N}
\label{eq3}
\end{equation}%
We use the subscript "support" because this coordination number does not
take into account the existence of non-supporting contacts. In
non-sequentially deposited beds there will be some contacts that do not
serve to the stability of any of the two touching particles. In our
simulations, these non-supporting contacts represent least than $0.3\mathbf{%
\%}$ of the contacts. Using Eqs. (\ref{eq1})--(\ref{eq3}) , and after some
algebra we obtain

\begin{equation}
\langle z\rangle _{\text{support}}=2\left[ 1+\frac{1}{N}\sum_{s=1}^{N}n(s)%
\right]  \label{eq4}
\end{equation}

The analysis above can be done also in 3D but it requires more complex
expressions due to the fact that arches can present branches \cite%
{Pugnaloni1,Pugnaloni2,Mehta1}.

Bearing in mind that $\sum_{s=1}^{N}sn(s)=N$, we can model $n(s)$ as a
simple exponential decay in the limit $N\rightarrow \infty $, i.e.,

\begin{equation}
n(s)/N=2(\cosh [\alpha ]-1)\exp [-\alpha s],  \label{eq5}
\end{equation}%
with $\alpha $ a positive number. In this model we obtain

\begin{equation}
\langle z\rangle _{\text{support}}=4-2\exp [-\alpha ].  \label{eq6}
\end{equation}

The exponential model seems to work well for string-like bridges in 3D (see
Ref. \cite{Mehta1}). We will see below that, in 2D, this is only a rough
model for $n(s)$ according to our simulation results. However, from Eq. \ref%
{eq6} we can already observe that $\langle z\rangle _{\text{support}}$ can
only vary between 2 and 4 in 2D. Notice that this result does not depend on
the shape of the particles so long as they are convex and support each other
through point contacts. The case $\langle z\rangle _{\text{support}}=2$
corresponds to a very particular configuration where all particles belong to
a single large arch. This, of course, is impossible to achieve with walls
present.

In a graph of $p_{2}/N$ as a function of $p_{1}/N$ (see Fig. 1) any packing
must belong to the triangle $[(0,0),(0,1),(1,0)]$ and the lines parallel to
the segment $[(0,0.5),(1,0)]$ correspond to states of equal $\langle
z\rangle _{\text{support}}$. The curve shown in Fig. 1 corresponds to the
loci of the states given by the exponential model of $n(s)$ with $\alpha $
ranging from $0$ to $\infty $. In practice, $\langle z\rangle _{\text{support%
}}$ only ranges from 3 to 4 (see the simulation data point in Fig. 1). The
lower limit ($\langle z\rangle _{\text{support}}=3$) has been suggested to
correspond to the marginally rigid state in 2D \cite{Blumenfeld1}.

\section{Simulation details}

Our simulations are based on an algorithm for inelastic hard disks designed
by Manna and Khakhar \cite{Manna1, Manna2}. This is a pseudo dynamics that
consists in small falls and rolls of the grains until they come to rest by
contacting other particles or the system boundaries. We use a container
formed by a flat base and two flat vertical walls. No periodic boundary
conditions are applied. Once all the grains come to rest, the system is
expanded and randomly shaken to simulate a vertical tap. Then, a new
deposition cycle begins. After several taps, the system achieves a steady
state where all characterizing parameters fluctuate around an equilibrium
value.

The deposition algorithm consists in picking up a disk in the system and
performing a free fall of length $\delta $ if the disk has no supporting
contacts, or a roll of arc-length $\delta $ over its supporting disk if the
disk has one single supporting contact \cite{Manna1,Manna2}. Disks with two
supporting contacts are considered stable and left in their positions. If in
the course of a fall of length $\delta $ a disk collides with another disk
(or the base), the falling disk is put just in contact and this contact is
defined as its \textit{first supporting contact}. Analogously, If in the
course of a roll of length $\delta $ a disk collides with another disk (or a
wall), the rolling disk is put just in contact. If the \textit{first
supporting contact} and the second contact are such that the disk is in a
stable position, the second contact is defined as the \textit{second
supporting contact}; otherwise, the lowest of the two contacting particle is
taken as the \textit{first supporting contact} of the rolling disk and the 
\textit{second supporting contact} is left undefined. If, during a roll, a
particle reaches a lower position than the supporting particle over which it
is rolling, its \textit{first supporting contact} is left undefined. A
moving disk can change the stability state of other disks supported by it,
therefore, this information is updated after each move. The deposition is
over once each particle in the system has both supporting contacts defined
or is in contact with the base (particles at the base are supported by a
single contact). Then, the coordinates of the centers of the disks and the
corresponding labels of the two supporting particles, wall, or base, are
saved for analysis.

The tapping of the system is simulated by multiplying the vertical
coordinate of each particle by $A$ (with $A>1$). Then, the particles are
subjected to several (about $20$) Monte Carlo loops where positions are
changed by displacing particles a random length $\Delta r$ uniformly
distributed in the range $0<\Delta r<A-1$. New configurations that
correspond to overlaps are rejected. This disordering phase is crucial to
avoid particles falling back again into the same positions. Moreover, the
upper limit for $\Delta r$ (i.e. $A-1$) is deliberately chosen so that a
larger tap promotes larger random changes in the particle positions.

The simulations are carried out in a rectangular box of width $20$
containing $2000$ equal-sized disks of radius $r=0.1+1/\sqrt{2}\approx 0.807$%
. The disk size is chosen in such way that the neighbor linked-cell system
used to speed up the simulation contains a single particle per cell.
Initially, disks are placed at random in the simulation box (with no
overlaps) and deposited using the pseudo dynamic algorithm. Then, $10^{3}$
tapping cycles are performed for equilibration followed by $10^{3}$ taps for
production; saving only $1$ every $10$ fully deposited configurations.
Tapping amplitudes range from $1.02$ up to $2.0$. Enlarging the width of the
simulation box up to $30$ has shown no effect on the results that we present
here. The same is true if one reduces the number of particles down to $500$.

An important point in this simulations is the effect that the parameter $%
\delta $ has in the results since particles do not move simultaneously but
one at a time. One might expect that in the limit $\delta \rightarrow 0$ we
should recover a fairly "realistic" dynamics for fully inelastic rough disk
dragged downwards at constant velocity. This should represent particles
deposited in a viscous medium or carried by a conveyor belt. In Fig. 2, we
show $\phi $ and $\langle z\rangle _{\text{support}}$ for $A=1.1$ as a
function of $\delta $. As we can see, the results are independent of $\delta 
$ for small values of the parameter. Since computation efficiency decreases
with decreasing $\delta $ ---due to the number of free falls required for
the particles to come together at the bottom of the container---, we choose
the largest value that yield results indistinguishable from the small-$%
\delta $ limit within statistical errors (i.e., $\delta =0.01$). This value
might be inappropriate for simulations with small values of $A$. We have
checked that results are consistent in such simulations by reducing $\delta $
up to an order of magnitude.

The deposited configurations are analyzed in search of arches. We first
identify all mutually stable particles ---which we define as directly
connected--- and then we find the arches as chains of connected particles.
Two disks A and B are mutually stable if A is the left supporting particle
of B and B is the right supporting particle of A, or viceversa. We measure
the total number of arches, arch size distribution $n(s)$, and the
horizontal span distribution of the arches $n_{s}(x)$. The latter is the
probability density of finding an arch consisting of $s$ disks with
horizontal span between $x$ and $x+dx$. The horizontal span (or lateral
extension) is defined as the projection onto the horizontal axis of the
segment that joins the centers of the right-end disk and the left-end disk
in the arch.

\section{Results}

In Fig. 3 we show the area fraction (or packing fraction) $\phi $ occupied
by the disks as a function of the tapping amplitude. We also plot in Fig. 3 $%
\langle z\rangle _{\text{support}}$ as a function of $A$. We can identify
four parts in Fig. 3(a). For small values of $A$ (up to about $1.1$), the
area fraction falls very slowly from just above $0.84$ to around $0.83$. In
the range $1.1<A<1.15$, simulations present large fluctuations but a sharp
drop is observed; with $\phi $ ranging between $0.76$ and $0.83$. Then, in
the range $1.15<A<1.5$ there is a mild decrease in $\phi $. Finally, for $%
A>1.5$, we can observe in Fig. 3(a) a very slow rise in $\phi $.

The sharp change in $\phi $ around $A=1.13$ has been described previously as
an order--disorder transition due to the crystallization of the disks into a
triangular lattice. This transition has been located around $\phi =0.8$ by
experiments \cite{Rankenburg1}, which agrees with our simulations. The
maximum value of $\phi $ in the crystalline regime depends strongly on the
ratio between the box width and the diameter of the disks. For integer
ratios, the crystal can achieve the smallest lattice constant, hence the
highest packing fraction. We do not commensurate our box to the particle
size since this situation is difficult to achieve in experiments anyway.

The slow increase in $\phi $ for very large tapping amplitudes is consistent
with experiments carried out using non-circular particles "deposited" by a
conveyor belt in a $\sqcup $-shaped container \cite{Blumenfeld1}. In this
regime, we expect $\phi $ to slowly approach the value of sequentially
deposited disks (i.e., $\phi \approxeq 0.82$ \cite{Barker2}), since a very
strong tap should separate out the disks to such degree that they will fall
back without many multi-particle collisions.

In Fig. 4 we show two examples of packings with $250$ particles
corresponding to the disordered [Fig. 4(a)] and the ordered [Fig. 4(b)]
regimes. Arches are indicated with segments that join mutually stable
particles. It is important to notice here that in our simulations particles
that reach the base "stick" to their positions and are not pushed aside by
further colliding disks during deposition. This prevents the formation of
completely regular crystals since the first layer is already a randomly
deposited layer. In a proper granular dynamic simulation one obtains more
ordered structures \cite{Arevalo1}. Our ordered regime is better described
as a layering of the system rather than a crystallization. This is indeed a
limitation of the model since most 2D granular beds tend to present
crystal-like order in real life.

It is interesting to see whether in the transition region ($1.1<A<1.15$) the
system presents a phase separation type behavior. We have plotted the
profile of $\phi $ in the vertical ($y$) direction for some tapping
amplitudes in Fig. 5. In the disordered regime, with exception of several
bottom layers, the system shows a constant $\phi (y)$ (with values below $%
0.8 $). In the ordered regime, the system has also a constant $\phi (y)$
(with values around $0.84$). Within the transition region, however, the
profile shows a monotonic decrease from ordered-like area fractions down to
disordered-like densities. We have found no sing of a step-wise profile
which would indicate a coexistence between an ordered phase and a disordered
phase.

The most striking results from our simulations is the behavior of $\langle
z\rangle _{\text{support}}$ as a function of $A$ [see Fig. 3(b)]. Here, we
can identify only three distinct parts. For small $A$ (within the ordered
region), $\langle z\rangle _{\text{support}}$ grows with $A$. This implies
that larger tapping amplitudes \ "destroy" mutual contacts in the ordered
phase. Interestingly, at the order--disorder transition, $\langle z\rangle _{%
\text{support}}$ presents a sharp drop. This may be attributed to the
appearance of many arches in the system that lead to the loss of order.
Finally, within the disordered region, the system increases its coordination
linearly with $A$. This last observation is consistent with experiments in
2D of particles "deposited" by a conveyor belt \cite{Blumenfeld1}. Clearly,
arches are always present in the system since $\langle z\rangle _{\text{%
support}}<4$ in all cases. The striking feature is the non-monotonic
behavior of $\langle z\rangle _{\text{support}}$. Starting from small
tapping amplitudes one can remove mutual contacts (and hence arches) by
increasing $A$. However, at the order--disorder transition a small increase
in $A$ leads to the creation of many new mutual contacts. Once in the new
disordered regime, one can again remove mutual contacts by a further
increase in $A$. Unfortunately, we are unable to support these findings with
experimental evidence (apart from the behavior in the disordered regime that
agrees with Blumenfeld et al. \cite{Blumenfeld1}) since coordination numbers
are rarely obtained in experiments. It is worth pointing out that
simulations of 3D systems show a monotonic \textit{decrease} (not increase!)
in $\langle z\rangle $ as $A$ is increased (see for example Ref. \cite%
{Mehta2}).

Form Eq. \ref{eq4} we know that the number of arches in the system is
inverse to $\langle z\rangle _{\text{support}}$, this is shown in Fig. 6
where we plot the total number of arches per particle. Of course, we see
again a sudden change at the order--disorder transition. Arches are more
rare in the ordered phase just before the transition to the disordered
phase. As we mentioned, in both regimes (ordered and disordered), arches are
"destroyed" by increasing the tapping amplitude.

It is commonly said that arches are responsible for voids in a granular pack
which, in turn, diminish $\phi $. However, from our results we can conclude
that, in 2D, this is only true within the order--disorder transition region (%
$1.1<A<1.15$)\ and for very large tapping amplitudes ($A>1.5$). In both
cases a decrease in the number of arches correspond to an increase in the
packing fraction, and viceversa. In contrast, for the rest of the packings
(i.e., for $A<1.1$ and $1.15<A<1.5$), a decrease in the number of arches
coincides with a decrease in packing fraction. This result, may seem rather
counterintuitive; however, is not just the number of arches but their
geometrical properties that determine the total volume of the voids left
beneath them. Let us just remind ourselves that random packings of disks
containing no arches at all (i.e., $\langle z\rangle _{\text{support}}=4$)
have $\phi \approxeq 0.82$ \cite{Barker2}, which is lower than our
arch-containing ordered packings and higher that our arch-containing
disordered packings.

In the inset of Fig. 6 we show the arch size distribution $n(s)$ for a
single tapping amplitude. This distribution is rather insensitive to $A$.
However, very small changes in $n(s)$ promote significant changes in $%
\langle z\rangle _{\text{support}}$. The arch size distribution can be
fitted very well to a parabola in a semilogarithmic plot. However, this
fitting can only be extended up to $s=7$ or $8$, since we find no arches
with more than $8$ disks in our simulations. Presumably, simulations with a
larger number of particles and a wider container will occasionally present
larger arches. From Fig. 6, we can see that the simple exponential model we
presented above is not suitable to describe arch distributions in 2D
packings.

We have also analyzed the horizontal span of the arches. In Fig. 7 we
represent $n_{s}(x)$ for $s=2$ (a)$,$ $3$ (b) and $4$ (c)$,$ and the
horizontal span $n(x)$ averaged over all arches [Fig. 7(d)]. We include
results from two simulations just before and after the order--disorder
transition along with the theoretical model presented by To et al. \cite{To1}
based on a restricted random walk. In the case of $n(x)$ we have no
theoretical prediction since a theoretical model for $n(s)$ is needed to
weight the contributions of each size of arch.

In the ordered regime, we can appreciate a clear discretization of the arch
extensions. This is easily understood since in the ordered state particles
---even those forming arches--- are organized in layers. Arches consisting
of two disks, for example, can be formed by two disks at the same layer
[corresponding to the peak near $x=1.0$ in Fig. 7(a)], or by one disk in one
layer and another disk in the next layer displaced half "lattice constant"
in $x$ (corresponding to the peak near $x=0.5$). This argument can be
extended to larger arches. See Fig. 4(b) to appreciate the form of the
arches in the ordered regime.

In the disordered regime, arches have no distinct lateral extensions but a
continuous distribution. This is in agreement with the simple model by To et
al. which was actually designed to represent arches at the outlet of a
hopper. However, our arches tend to be more extended than those from the
restricted random walk model. In particular, for two-disk-arches, we have no
incidence of zero lateral span (corresponding to one disk on top of the
other) in contrast with the model.

Finally, in order to check the ability of the simulation model to reproduce
some features typical of granular materials, we have also performed an
annealed tapping on the system. We have increased and decreased $A$
progressively (from $1.0$ to $2.0$) five times and have plotted $\phi $ and $%
\langle z\rangle _{\text{support}}$ as a function of $A$ averaging all $A$%
-increasing cycles and all $A$-decreasing cycles separately. The very first $%
A$-increasing cycle is kept aside. Each increase from $A=1.0$ to $A=2.0$
takes $2000$ taps. The results in Fig. 8 correspond to averages over 5
independent simulation using $500$ particles. For reference, we include the
steady state curves obtained at constant tapping amplitude from Fig. 3. The
inset shows results where the ramp rate is increased so that each cycle
takes only $200$ taps instead of $2000$.

According to Nowak et al. \cite{Nowak1}, after the very first increasing
cycle, a granular bed should enter a reversible regime where further cycles
of the tapping amplitude follow the same curve in the $\phi $--$A$ plot.
However, Mehta and Barker \cite{Barker1} found in their 3D simulations a
hysteresis loop and suggest that the area of the loop should increase as the
ramp rate is increased. We also find hysteresis in our 2D simulations.
However, increasing the ramp rate does not clearly enlarge the hysteresis
loop. The most clear change due to the increase in ramp rate is the
difficulty that the system finds in reaching the ordered regime. This is in
agreement with the experiments by Nowak et al. where an increase in the ramp
rate leads to lower densities in the small-$A$ limit.

\section{Conclusions}

We have shown through a pseudo dynamic simulation that arches are ubiquitous
in a 2D granular packing. Very light tapping of the granular sample promotes
layering of the particles. The transition from large-amplitude tapping
(disordered packings) to small-amplitude tapping (ordered packings) is very
sharp and occurs without a phase separation mechanism. Interestingly, we
found that in a wide range of tapping amplitudes an increase in the number
of arches coincides with an increase of the density of the packing. With
exception of the order--disorder transition region, the mean coordination
number is always increased by increasing the tapping amplitude; in contrast
to what is found in 3D packings.

\begin{acknowledgments}
This work has been supported by CONICET of Argentina. M.G.V. and L.G.V. are
fellows of CONICET. L.A.P. is a member of CONICET.
\end{acknowledgments}

\begin{center}
\textbf{Figure captions}
\end{center}

\bigskip

FIG. 1. Mutual stability phase diagram for 2D packings. Here, $p_{1}/N$ and $%
p_{2}/N$ are the fractions of particles that present 1 and 2 mutually stable
supporting contacts, respectively. The state point of any packing must lie
in the triangle defined by ($0,0$), ($0,1$) and ($1,0$). The dotted lines
correspond to states of equal $\langle z\rangle _{\text{support}}$ (the
numbers indicate the corresponding values). The solid curve represents all
the states that can be obtained through the exponential model (see text).
Open circles correspond to a representative sample of the packings generated
in our simulations.

\bigskip

FIG. 2. Packing fraction (a) and $\langle z\rangle _{\text{support}}$ (b) as
a function of the parameter $\delta $ of the simulation algorithm for $A=1.1$%
. Solid lines are only to guide the eye.

\bigskip

FIG. 3. Packing fraction (a) and $\langle z\rangle _{\text{support}}$ (b) as
a function of the tapping amplitude $A$.

\bigskip

FIG. 4. Examples of the packings. (a) $250$ particles tapped with $A=1.3$
which corresponds to the disordered regime. (b) $250$ particles tapped with $%
A=1.1$ which corresponds to the ordered regime. Arches are indicated by
segments joining mutually stable particles.

\bigskip

FIG. 5. Vertical profile of the packing fraction for various values of $A$:
squares ($1.1$), circles ($1.13$), up triangles ($1.135$), stars ($1.14$),
diamonds ($1.15$), down triangles ($1.25$), and pentagons ($2.0$).

\bigskip

FIG. 6. Number of arches per unit particle as a function of the tapping
amplitude. The solid line is only a guide to the eye. The inset shows a
semilog plot of the arch size distribution $n(s)$. Symbols correspond to our
simulations for $A=1.3$ and the dotted line to a fit with a second order
polynomial.

\bigskip

FIG. 7. Distribution of the horizontal span $n_{s}(x)$ of the arches: (a)
arches with two disks, (b) arches with three disks, and (c) arches with four
disks. Part (d) corresponds to the horizontal span distribution over all
arches. The dotted line corresponds to $A=1.2$. The dashed line corresponds
to $A=1.1$. Solid line corresponds to the restricted random walk model (see
Ref. \cite{To1}).

\bigskip

FIG. 8. Packing fraction (a) and $\langle z\rangle _{\text{support}}$ (b) as
a function of the tapping amplitude $A$ along an annealed tapping. Tapping
amplitude is increased and decreased several cycles. Each increasing
(decreasing) cycle takes $2000$ taps. As a reference, we show the steady
state obtained through constant tapping amplitude with symbols and solid
thick line (see Fig. 3). The solid thin line corresponds to the initial
increase of $A$. The dashed line corresponds to the decreasing phase and the
dotted line to the increasing phase. All increasing cycles (apart from the
initial increase) are averaged to produce a single curve. The same is done
with the decreasing cycles. The insets show results where each increasing
(decreasing) cycle is performed in only $200$ taps.

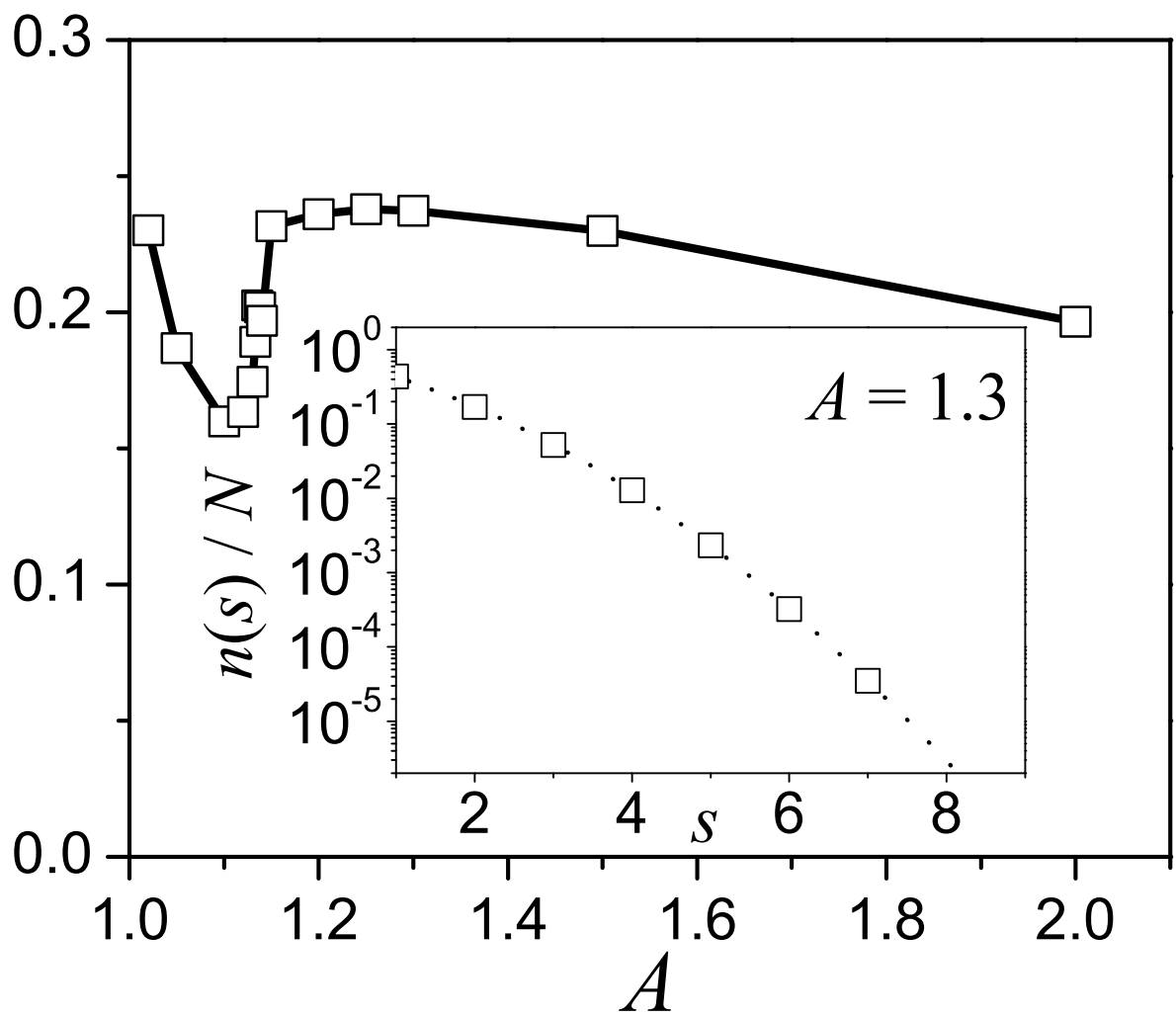
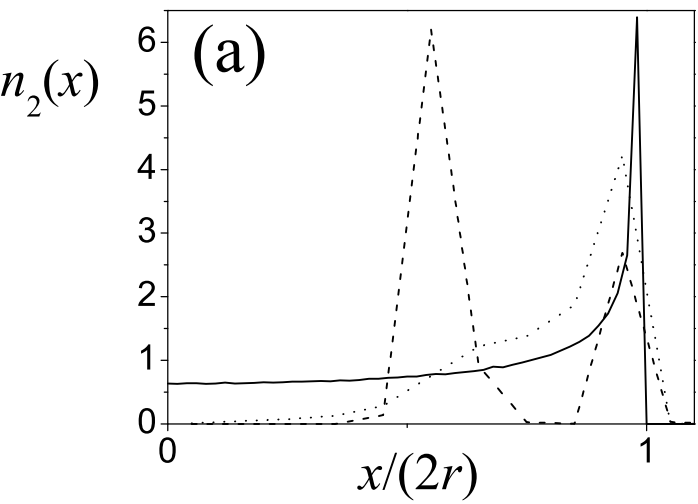
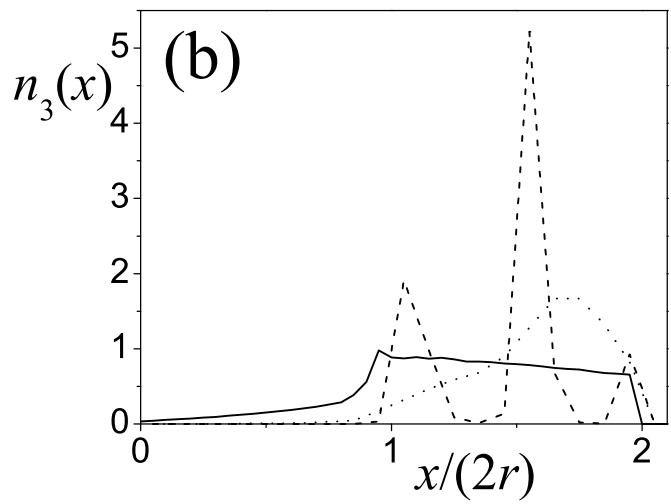
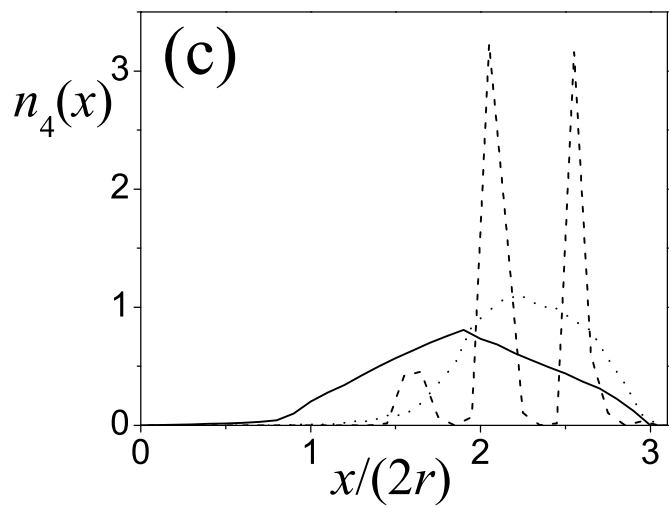
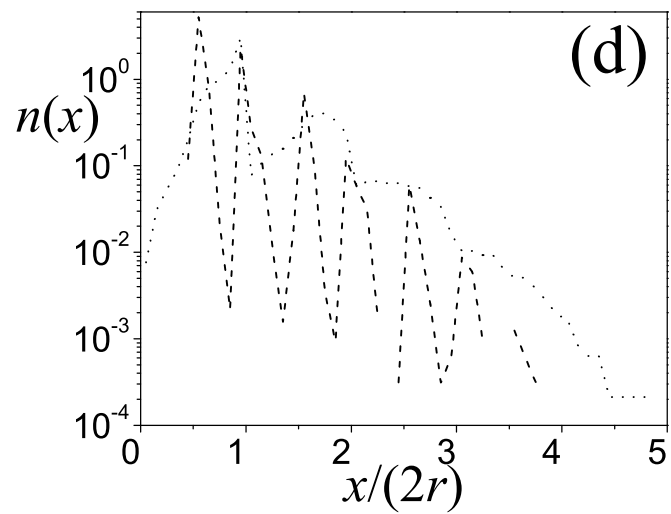
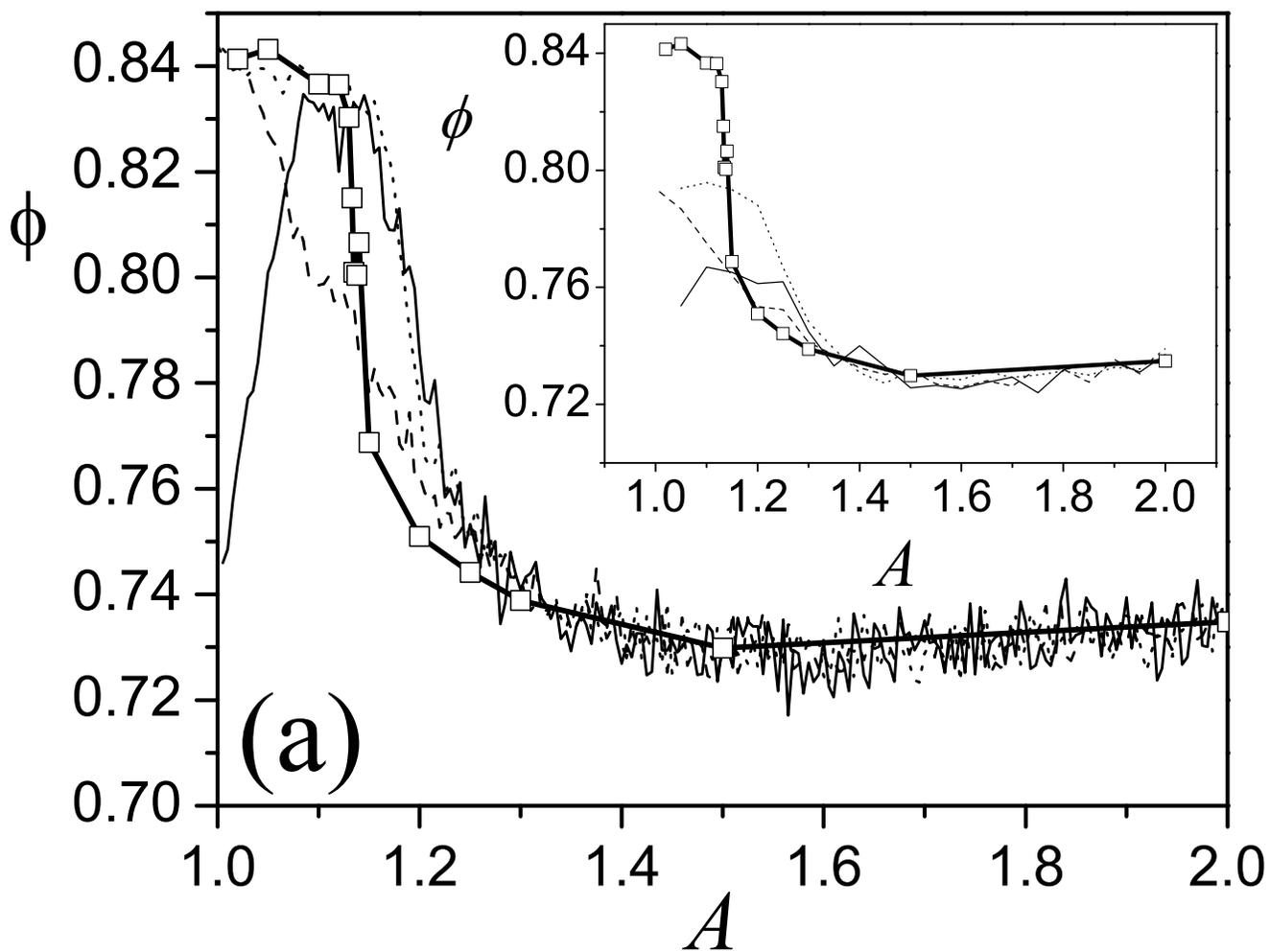
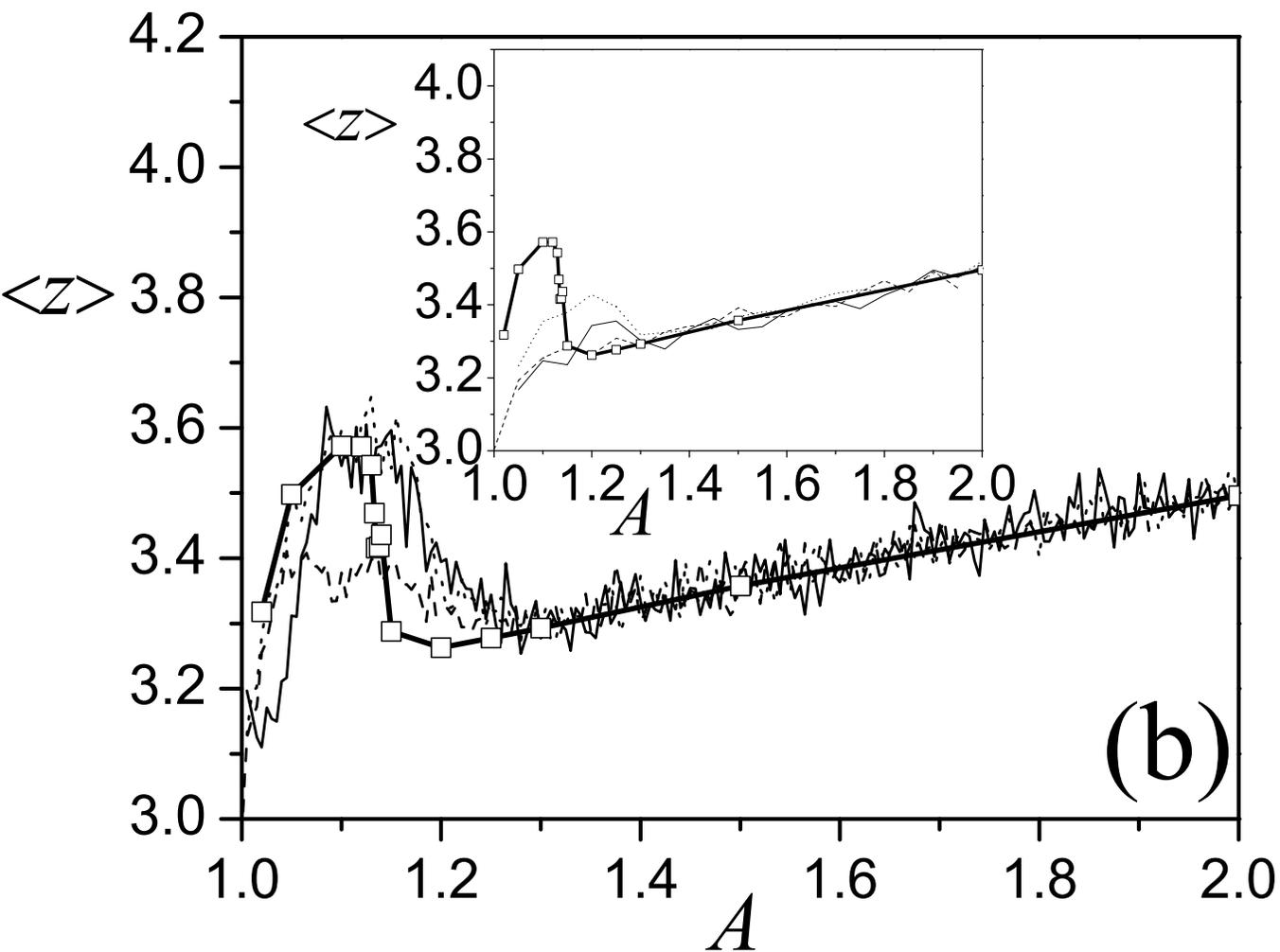
\end{document}